\documentclass[final,times]{elsarticle}
\pdfoutput=1

\usepackage{graphicx}
\usepackage{amsmath,amsfonts,amssymb}

\usepackage{lineno}
\usepackage{hyperref}

\begin{document}

\begin{frontmatter}


\title{A quantitative definition of organismality and its application to lichen}



\author[a,d]{Eric Libby}
\author[a,b]{Joshua A. Grochow} 
\author[c,a]{Simon DeDeo} 
\author[a]{David Wolpert}

\address[a]{Santa Fe Institute, 1399 Hyde Park Road, Santa Fe, NM 87501, USA}
\address[b]{Depts. of Computer Science and Mathematics, University of Colorado at Boulder, 1111 Engineering Drive, ECOT 430 UCB, Boulder, CO 80309, USA}
\address[c]{Dept. of Social and Decision Sciences, Carnegie Mellon University, 5000 Forbes Avenue, BP 208, Pittsburgh, PA 15213, USA}
\address[d]{corresponding author email: elibby@santafe.edu}

\begin{abstract}
The organism is a fundamental concept in biology. However there is no universally accepted, formal, and yet broadly applicable definition of what an organism is. Here we introduce a candidate definition. We adopt the view that the ``organism'' is a \emph{functional} concept, used by scientists to address particular questions concerning the future state of a biological system, rather than something wholly defined by that system. In this approach organisms are a coarse-graining of a fine-grained dynamical model of a biological system. Crucially, the coarse-graining of the system into organisms is chosen so that their dynamics can be used by scientists to make accurate predictions of \emph{those features of the biological system that interests them}, and do so with minimal computational burden. To illustrate our framework we apply it to a dynamic model of lichen symbiosis---a system where either the lichen or its constituent fungi and algae could reasonably be considered ``organisms.'' We find that the best choice for what organisms are in this scenario are complex mixtures of many entities that do not resemble standard notions of organisms. When we restrict our allowed coarse-grainings to more traditional types of organisms, we find that ecological conditions, such as niche competition and predation pressure, play a significant role in determining the best choice for organisms. 
\end{abstract}

\begin{keyword}
organismality \sep state space compression \sep lichen \sep individuality

\end{keyword}

\end{frontmatter}

\section*{Introduction}
The organism is a fundamental unit of biology, key to a vast range of biological theory. Therefore it is important to have a formal and broadly applicable definition of ``organism'' that identifies the organisms in any given biological system. However despite its foundational role, there is little consensus on what is and is not an organism in any broad sense \cite{BENSON:1989ex,Wolfe,Nicholson:2014in,LAUBICHLER:2000bh}.  Beyond navigating the difficulty of what it means to be alive, one key challenge is distinguishing between parts of organisms, organisms, and groups of organisms \cite{Pepper:2008gk}. In some cases, the distinction seems fairly unequivocal. For example, a liver is part of an organism, a human is an organism, and a nation is a group of organisms. However, there are many other instances in which the distinction is more nebulous. For example, in a grove of quaking aspens {\it Populus tremuloides} hundreds of trees share a common root system. Is each tree an organism or is the whole grove? Many other problematic examples exist \cite{godfrey2009darwinian}. This has driven some to embrace the ambiguity, and consider a spectrum of organismality, rendering the concept somewhat meaningless \cite{Queller:2009gl}. 
\par
As an alternative, here we adopt the position of ``working biologists'' \cite{Pepper:2008gk}, using it as the basis for defining organisms. Specifically, we introduce a novel approach that determines the ``organisms'' in a given biological system \emph{in the context of the predictions concerning the system that the scientist wishes to make}, in addition to the specific nature of the system.
We focus on what the concept of an organism is \emph{for}, viewing it as a mathematical construct that allows us to predict the future dynamics of certain quantities of interest.
\par
From this perspective, the ``organism'' is not an inherent feature of a biological system but depends on the scientific question being asked and how best to model the system in order to answer that question. For example, when studying human migration patterns, a human's microbiome might reasonably be considered as part of the human, but when studying human disease, the microbiome might better be seen as a community of microorganisms each of which is an organism. The choice depends on what aspect of the system interests the scientist. 
\par
This approach avoids the difficulties that have plagued earlier attempts to formulate a universally applicable definition of the organism solely in terms of attributes of a system under study. For example, while functional integration is important to some \cite{DavidSloanWilson:2009vu} of these attempts, it is intentionally omitted by others \cite{Clarke:2010ff} as being inappropriate in many domains. Even features that might seem fundamental to the notion of an organism are problematic. For example reproduction, in the broad sense, is difficult to define in a way that distinguishes it from growth \cite{godfrey2009darwinian}. Similarly, specific forms of reproduction, say via a dedicated germ line or via sex, are construed by some as an indicator of organisms \cite{Clarke:2010ff}. However these approaches exclude unicellular taxa that reproduce asexually.   
\par
This approach also explicitly allows for the possibility that no single specification of the organism in a given biological system is suitable for predicting the future dynamics of all quantities of interest concerning that system; an entity may be an ``organism'' in the context of predicting one quantity but only part of an ``organism'' when predicting a different one. This approach purposely has nothing to say about the ontological status of an organism. It does, however, accord with the common practice of scientists.
\par
Here we show how to formalize this approach, using State Space Compression (SSC) \cite{Wolpert2017}. In SSC there are two criteria for specifying the organisms in a dynamic system, 
with the final choice of how to specify the organisms determined by trading those criteria off one another. First, it must be possible to make accurate predictions of future values of \emph{the exogenously specified quantity the scientist is interested in} by considering the dynamics of those individuals. Second, making those predictions must require as little computation as possible.  
We show how to use these two criteria to specify the organisms in a biological
system. We then apply it to a biological system with different potential choices for how to
specify the organism: a dynamic system of interacting lichens, fungi, and algae. 
In particular we show how natural variations of our model change the choice of which
specification of the organism is best and how that relationship depends 
on the quantity being predicted.

\subsection*{Our Framework: State Space Compression}
\begin{figure}
\centering
\includegraphics[width=.8\linewidth]{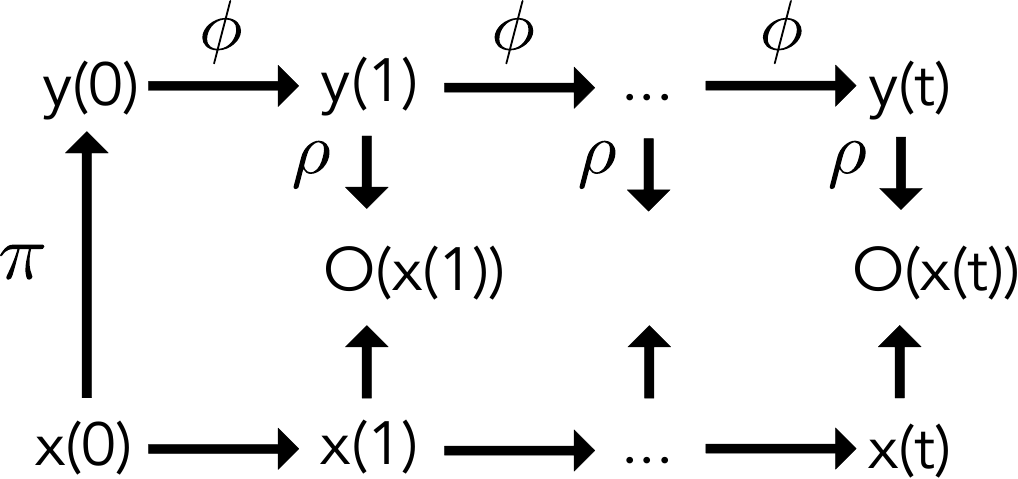}
\caption{{\bf State Space Compression schematic.} The microstate variables $x$ are compressed at $t=0$ into a set of macrostate variables $y$ via the $\pi$ map. The dynamics of the macrostate variables are governed by the map $\phi$. At each time step $\rho$ maps the macrostate variables into some quantity of interest in space $O$.}
\label{fig:FigSSC}
\end{figure}
In SSC we identify organisms as coarse-grained states of a system which can be used to predict a future value of some quantity of interest. 
This reflects the view implicit in \emph{any} analysis of a dynamically evolving biological system
that reduces it to a set of interacting individuals: 
although the system may be high-dimensional, the quantities of interest can be well-approximated by dynamics in many fewer dimensions.

\par
Formally, in SSC we start with some model of the dynamical system with potentially high-dimensional
state space $X$ (called the ``microstate space''). We also assume we have an exogenously provided
map taking any $x \in X$ into the quantities of interest, which live in some space $O$ (for ``observable of interest''). We define a 
\textbf{state space compression}
of such a dynamical system as a set of three  maps (see Figure \ref{fig:FigSSC} for a schematic):  

1) A compression map $\pi$ that coarse-grains initial microstates $x \in X$ into initial \textbf{macrostates} (which typically live in a lower dimensional macrospace $Y$).

2) A map $\phi$ that governs the dynamics of the macrostates;

3) A map $\rho$ that converts macrostates into the space $O$. 

\noindent Importantly, no new information flows up from $x$ to affect the dynamics of $y$ once the initial $y$ has been set. This is an important distinction between SSC and other formalizations of ``macrostates'' that have been proposed in the literature \cite{Pfante:2014dt,AY:2008cc,Shalizi:2003ul}.
\par
The quality of such a state-space compression is judged by the trade-off it achieves between accuracy of prediction of the quantity of interest and the computational (as well as measurement and other related) cost needed to make that prediction. Thus SSC also needs two cost functions to be specified: 

1) A function that measures the predictive accuracy of an SSC set of maps, i.\,e., how well $\rho(\phi^t(\pi(x(0)))$ approximates $O(x(t))$, where $\phi^t$ corresponds to $t$ iterations of the $\phi$ map;

2) A function that measures the computational cost of running an
SSC set of maps, $\pi \rightarrow \phi^t \rightarrow \rho$. This cost will often reward compressions into a lower-dimensional coarse-grained space, but may reward other computationally useful features as well, such as sparsity or other structures. 

If accuracy were the sole criterion for determining the best compression, then the identity ``compression'' that
simply maps $x$ to itself would be optimal. It is precisely when there are compressions that (possibly greatly) reduce the computational burden of accurately predicting the quantity of interest that scientists use such compressions. 



In our illustration of SSC in this paper, the three maps $\pi, \phi$ and $\rho$ are all parameterized functions, so that the fitting process consists of finding optimal values of their parameters. There are many possible choices for such parameterized functions, and many possible algorithms for searching for the best associated values of the parameters. Here our focus is on a restricted class of parameterized functions that are easy to interpret, and ``make sense'' in a biological context. While we optimize over this class, we expect that in general the global optimum over \emph{all} functions lies outside this class.

There is a vast body of literature on generalized coarse-grainings, similar to those we consider via SSC. This includes work on causal states and computational mechanics \cite{crutchfieldReview, marzenCrutchfield}), state aggregation and lumpability \cite{auger2008aggregation, simon1961aggregation, chiappori2011new, white2000lumpable}, finite state projection \cite{munsky2008finite, munsky2006finite}, and reduced-order modeling \cite{moore1981PCA, lorenz, mori, zwanzig, holmesLumleyBerkooz, chorinHaldBook,antoulas, schilders2008model, beckLallLiangWest, antoulasSorenson, dengThesis, bai2002krylov,peherstorferLocal}. On the one hand, SSC differs from these works in that the computation cost of \emph{running the compressed dynamics} (as opposed to finding the optimal compression) is explicitly taken into account as part of our optimization; while most of the above works explicitly consider accuracy cost, and many of them explicitly consider the cost of computing the optimal compression, few if any of them explicitly consider the cost of running the compressed dynamics. SSC thus presents a great opportunity for adapting these methods to explicitly take into account the computation cost of the compressed model. Conversely, the plethora of techniques already available for compressing dynamics will surely be of use in further applications of SSC.



\subsection*{Lichens as a model system}
We illustrate our approach on a case study in organismality: the lichen symbiosis. Our goal in this paper is not to 
uncover any new understanding of the lichen symbiosis \emph{per se}, but rather to explore how SSC behaves in a representative biological domain.

Lichens are communities of fungi and photobionts (algae or cyanobacteria) that are capable of surviving in extreme environments. It can be argued that it is not appropriate to view any such community as a single organism because the fungi and photobionts have distinct genomes and are capable of living independent of the lichen association. Indeed, the relationship between them in a lichen has been described as a type of agricultural association in which the fungi are harvesting energy from the photobionts. Moreover, the lichen association is not a simple one-to-one relationship between fungus and photobiont species. Many lichens include another fungal partner \cite{Spribille:2016iv} or multiple types of photobiont (e.\,g., both cyanobacteria and algae). There can even be lichen associations of hundreds of different genomes \cite{Lucking:2014ee}. 
\par
On the other hand, separate lichens do exhibit a high degree of functional integration and can interact differently with their environment than do free-living, non-lichenized fungi or photobionts. For example, they can colonize unique niches and face predators that free-living fungi or photobionts do not due to their smaller size. In addition, even though lichens are composed of different genotypes, the collective can reproduce as a whole, making it a type of replicator that may be capable of Darwinian evolution. Indeed, collective reproduction has been identified as an important feature in reducing inter-species conflict and producing new individuals \cite{Kiers:2015eq}. It is this very ambiguity of how best to view lichens that makes them well-suited to the SSC approach.
\par
\subsection*{Microstate model}

\begin{figure}
\centering
\includegraphics[width=1\linewidth]{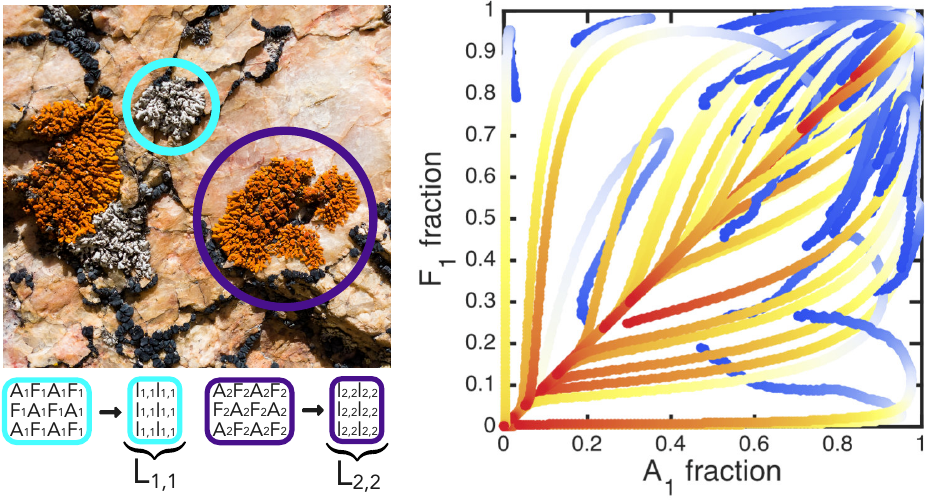}
\caption{{\bf Lichen microstate model.} (left) A photo of lichen communities in Santa Fe, NM, USA shows macro-level entities with defined boundaries. This might lead one to conclude two different lichen communities (indicated by circles) are each organisms in their own right. Our microstate model considers the dynamics of free-living fungi and algae and the lichens that they can form. Our model distinguishes between the lichen as a whole community (denoted by an uppercase L) and its constituent algal-fungal pairs (denoted by a lowercase $l$). By computing the dynamics of our microstate model and applying SSC we can determine the relevant organisms in terms of predicting quantities of interest. (right) Randomly chosen trajectories of the two quantities of interest---the fraction of $F_1$ and $A_1$ genotypes in the population---are shown plotted against one another. The coloring indicates the temporal aspect of the trajectories: time advances as blue moves to red. In our microstate model, the free-living $F_1$ and $A_1$ species are fitter than their counterpart free-living species, but they form a lichen that is less fit. Thus, in the short term they are more abundant but as the longer time scale lichen processes take hold they decrease in frequency.}
\label{fig:FigMicrostate}
\end{figure}
Our microstate model of lichen consists of 12 types of entities that interact in a spatially-explicit, discrete-time, agent-based simulation
 (see Methods for more details). There are two species of free-living algae ($A_1$ and $A_2$) and two species of free-living fungi ($F_1$ and $F_2$), all of which can survive and reproduce on their own. If a fungus and alga cell are close enough in space then they can form a lichen association, which happens with some specified probability. We assume that once a lichen is formed, all fungi and algae within that lichen reproduce at the same rate so that they maintain a constant 1-1 ratio. (In wild lichen populations this is unlikely to be the case, but such additional detail could easily be incorporated into the microspace model.) Since lichenized fungi and algae reproduce at the same rate it will sometimes be convenient to refer to lichenized alga-fungus pairs as lichen ``cells.'' 
 \par
 The four types of lichen cells are represented by the binary combinations of fungi and algae: $l_{1,1}$, $l_{1,2}$,$l_{2,1}$, \& $l_{2,2}$, where the first index is the alga and the second is the fungus, e.\,g., $l_{1,2}$ denotes a combination of $A_1$ and $F_2$. As lichen cells reproduce they contribute to the growth of a larger, multicellular lichen community. We denote lichen communities as $L_{1,1}$, $L_{1,2}$,$L_{2,1}$, \& $L_{2,2}$ where lichen cell $l_{i,j}$ forms community $L_{i,j}$. So a set of $m$ $l_{1,1}$ cells (alga-fungus pairs) can be partitioned into $k$ distinct $L_{1,1}$ lichen communities, where $k \leq m$. Figure~\ref{fig:FigMicrostate} describes the relationship between real lichen communities and lichen cells and their representations in our microstate model. 
 \par
With some fixed probability reproduction of a lichen cell results in the formation of a new lichen community, i.\,e., there is asexual reproduction of lichen communities. Thus, when a lichen cell reproduces it may either lead to the growth or reproduction of its lichen community. We track the allocation of lichen cells into lichen communities in our microstates, to allow our SSC approach to exploit this information in determining organismality. 
 
To introduce realistic and interesting contrast into this model, we stipulate that $F_1$ is fitter than $F_2$, and that $A_1$ is fitter than $A_2$, but that the lichen they form, $l_{1,1}$, is less fit than $l_{2,2}$. Since the free-living fungi and algae reproduce faster than their lichenized forms, the proportion of the population of $A_1$ and $F_1$ increases in the short term. But as lichens form and outcompete free-living species, the $l_{2,2}$ lichen cells take over the population because of their fitness advantage.

\subsection*{Compression map (SSC \texorpdfstring{$\pi$}{pi})}
Although our microstate model includes spatial information, we only consider compressions that ignore this spatial information; on the one hand, this represents an ansatz based on our intuitions about the quantities of interest, but it also has the important benefit of making \emph{solving} for the SSC-optimal compression much more feasible. The $\pi$ map thus compresses a 12-dimensional microstate variable $x$ into a macrospace variable $y$ of dimension $n$, $n \leq 12$. While in principle the $\pi$ map could take many functional forms, we only consider linear maps $a x = y$ where $a$ is an $n \times 12$ matrix. 
This choice reflects the assumption that organisms are composed of parts, just as a multicellular entity is composed of interacting single cells. 
Thus, the matrix $a$ describes how to combine components of $x$ into new macrostate individuals $y$:
\begin{eqnarray}
y_i & = &\underbrace{\big( a_{i,1} A_1 + a_{i,2} A_2 + a_{i,3} F_1 + a_{i,4} F_2 \big)}_{\text {\normalfont free-living algae and fungi}} \label{pimap} \\
\ & \ & + \underbrace{\big( a_{i,5} l_{1,1}+ a_{i,6} l_{1,2} + a_{i,7} l_{2,1}+ a_{i,8} l_{2,2} \big)}_{\text {\normalfont lichenized alga-fungus pairs}} \nonumber \\ 
\ & \ & +  \underbrace{\big( a_{i,9} L_{1,1}+ a_{i,10} L_{1,2} + a_{i,11} L_{2,1}+ a_{i,12} L_{2,2} \big)}_{\text {\normalfont lichen groups}} \nonumber
\end{eqnarray}

\subsection*{Macrostate dynamics (SSC \texorpdfstring{$\phi$}{phi})}
The $\phi$ map defines the dynamics of the macrostate variables. We choose $\phi$ to be discrete Lotka--Volterra equations,
\begin{equation}
\label{phimap}
y_i(t+1) = y_i(t) \Big( b_{i,0} + \sum_{j=1}^{n} b_{i,j} y_j \Big) {\text{\normalfont , for }}  i=1,\ldots,n 
\end{equation}
\noindent (The $b$ parameters in Eq.~\ref{phimap} correspond to linear growth terms and quadratic interaction terms that appear in the Lotka--Volterra equations.)
One advantage of this choice is that these equations are commonly used to simulate the dynamics of biological species, and are particularly easy to interpret. In addition, using these equations means that if any macrostate variable goes to zero then it stays at zero. This reflects the constraint that if any macrostate individual goes extinct then it cannot be regenerated from another individual.

\subsection*{Quantities of interest}
There are many natural choices for the quantities of interest, $O(x(t))$. Here, we choose quantities that may be of interest to population geneticists: the temporal dynamics of the population proportion of $A_1$ algal and $F_1$ fungal genotypes (referred to as $Q_1$ and $Q_2$ respectively).  Eq.~\ref{qoi} shows how we map the microstates into the two quantities of interest. Since lichen cells also include algal and fungal genomes, they are included in the proportion data. \begin{eqnarray}
Q_1  &=   \big( A_1+l_{1,1}+l_{1,2} \big) / \big(A_1+A_2+l_{1,1}+l_{1,2}+l_{2,1}+l_{2,2} \big) \label{qoi} \\
Q_2  &=  \big( F_1+l_{1,1}+l_{2,1} \big) / \big(F_1+F_2+l_{1,1}+l_{1,2}+l_{2,1}+l_{2,2} \big) \nonumber
\end{eqnarray}
Figure~\ref{fig:FigMicrostate} shows the relationship between these two quantities of interest through time resulting from the microspace dynamics. Both quantities of interest approach 1, making up 100\% of the population, but then decrease towards 0\% of the population. This reflects the choice that $A_1$ and $F_1$ are the fitter free-living species and $l_{2,2}$, composed of $A_2$ and $F_2$ genomes, is the fittest lichen cell. 

\subsection*{Prediction map (SSC \texorpdfstring{$\rho$}{rho})}
Since the quantities of interest are proportions of linear combinations of the microstate variables, we consider a complementary class of possible $\rho$ maps, each of which is given by a proportion of two linear combinations of the macrostate variables:
\begin{equation}
\label{rhomap}
\rho_k(y) = \Big( \sum_{j=1}^{n} c_{k,j} y_j \Big) / \Big( \sum_{j=1}^{n} d_{k,j} y_j \Big) {\text{\normalfont , for }}  k=1,2. 
\end{equation}
\noindent The $c$ and $d$ parameters in Eq.~\ref{rhomap} are coefficients of the proportion that must be fit in the process of optimizing the objective function, and $k=1,2$ indexes the two quantities of interest.

\section*{Results}
\subsection*{SSC maps that minimize accuracy cost}
\begin{figure}
\centering
\includegraphics[width=.8\linewidth]{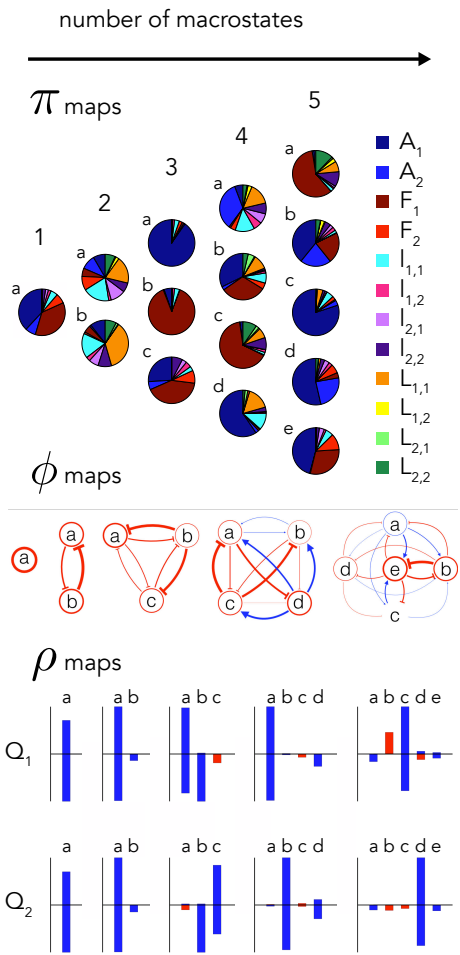}
\caption{{\bf The best fit maps for SSC that minimize accuracy cost}. The pie charts show how the $\pi$ maps combine the microstate variables into new macrostate individuals, each denoted by a lower case Roman letter (the ordering is arbitrary). The $\phi$ maps show the interactions between the macrostate individuals denoted by the same letters used for the pie charts that illustrate the $\pi$ maps. Red lines indicate negative interactions, blue lines indicate positive interactions, and the thickness of the line indicates the strength of the interaction corresponding to the fit coefficients in the Lotka--Volterra equations. The presence of a circle around the letter indicates whether there is a self interaction. The $\rho$ maps show how the macrostate variables are combined into terms in the numerator (bars above the black line) or the denominator (bars below the black line). The height of the bar determines the magnitude of the coefficient and the color determines whether it is positive (blue) or negative (red).}
\label{fig:Maps}
\end{figure}

We optimized the accuracy cost given by the triple of SSC maps $\pi, \phi$ and $\rho$ separately, for each number of macrostates (i.\,e., dimension) between 1 and 5 (see Figure~\ref{fig:Maps}). We performed those optimizations by fitting to the quantities of interest generated by 100 independent runs of the agent-based microstate model. The $\pi$ maps show that the macrostate variables, i.\,e., the ``organisms,'' are heterogeneous individuals that mix microstate variables. One general property that is shared across macrospaces is that the free-living species ($A_1, A_2, F_1, F_2$) make the largest contribution to the macrostate variables---the notable exceptions are individuals for the two-dimensional macrostate. 

The $\phi$ maps, which describe the interactions between the macrostate variables, show that the macrostate individuals interact in different ways as the size of the macrospace increases. For macrostate dimensions 1, 2 and 3 the graphs are fully connected and all interactions are inhibitory. In contrast, the graphs for dimensions 4 and 5 are not fully connected and include positive interactions between macrostate individuals. 

The $\rho$ maps for both the 1- and 2-dimensional macrospaces are similar for the two quantities of interest. This suggests that the fits are converging on the best average between the two quantities of interest and not distinguishing between them. When we checked the best 5 fits (out of 25), we find that all $\rho$ maps have this property. This contrasts with higher-dimensional macrospaces where there are distinct profiles for mapping between the macrostate variables and the quantities of interest.
\par
\begin{figure}
\centering
\includegraphics[width=1\linewidth]{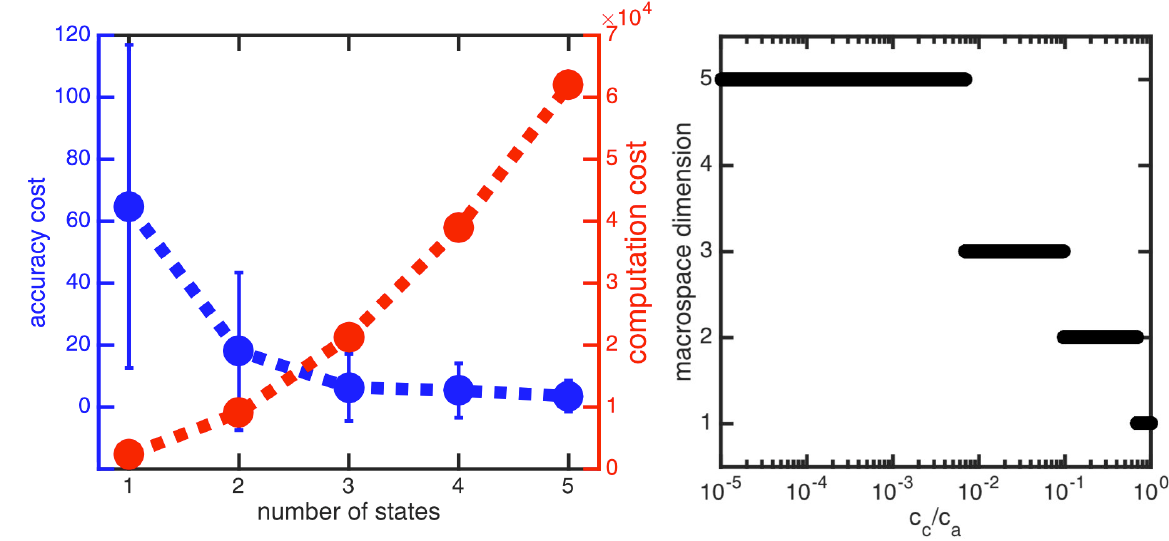}
\caption{{\bf The balance between accuracy and computation costs.} (left) The average accuracy cost on 50 new trajectories is shown as a function of macrospace dimension with error bars corresponding to the standard deviation. The computation cost is determined by the cost of the mathematical operations that need to be performed to evaluate the three maps on a single trajectory. (right) The optimal macrospace dimension is shown as a function of the ratio of coefficients that determine the contribution of computation and accuracy costs to the objective function.}
\label{fig:FigAccComp}
\end{figure}

While Figure~\ref{fig:Maps} shows the maps that minimize the accuracy cost for different macrospace sizes, it does not contribute to determining which macrospace is optimal. To do this we evaluated the accuracy of each collection of SSC maps for the different macrospace dimensions. The resulting accuracy costs are shown in Figure \ref{fig:FigAccComp}. The accuracy cost decreases monotonically with respect to increasing macrospace dimension; this accords with the fact that a higher-dimensional macrospace can reflect at least as much information as a lower-dimensional macrospace. In SSC, the optimal macrospace size depends on a balance between accuracy and computation costs.
\par
\par
\par
\subsection*{Computation cost}
To evaluate the cost of computation as a function of macrospace dimension, we determine the computation cost of each map in terms of the mathematical operations that need to be performed (see Methods: Determining the computation cost). Using this approach the computation cost of $\pi$,$\phi$,and $\rho$ is: 
\begin{equation}
O_x \big( 6 t n^2 + n(23-3 t)+ 2t \big),
 \label{Eq:totcostquick}
\end{equation}
where $O_x$ is the cost of a single scalar addition, division, or multiplication step, $n$ is the macrospace dimension, and $t$ is the length of the dynamical trajectory. Eq.~\ref{Eq:totcostquick} represents one approach to assigning a cost to computation, but there are many other ways (see examples in \cite{Wolpert2017}). The choice ultimately depends on the needs and equipment of the scientist.  

\par
\subsection*{Balancing accuracy and computation costs}
In this paper, we assume that the best compression strikes a balance between the cost of making an inaccurate prediction and the cost to run the model (the computation cost). Although there can be many ways to do this, we assume that the scientist must minimize an objective function $f(n)$ that is a linear combination of accuracy cost ($A(n)$) and computation cost $C(n)$.
\begin{equation}
f(n) = c_a A(n) + c_c C(n)
\end{equation}
The coefficients $c_a$ and $c_c$ assign the weight of each of the costs.  When comparing macrospace dimensions, we see that for any two $n,n'$, $f(n) \leq f(n')$ if and only if $A(n) + \frac{c_c}{c_a} C(n) \leq A(n') + \frac{c_c}{c_a} C(n)$. 
We can therefore express the best macrostate as a function of the ratio $\frac{c_c}{c_a}$ (see Figure~\ref{fig:FigAccComp}). 
The optimal macrostate dimension moves from $5$ states (the maximum we considered) when improving accuracy is more important to $1$ state when computation is costly. We note that this progression skips the 4-dimensional macrospace. This occurs because the accuracy cost drops the least from dimensions 3 to 4 while the computation cost increases quadratically.
\par
\subsection*{Measurement cost}
\begin{figure}
\centering
\includegraphics[width=.5\linewidth]{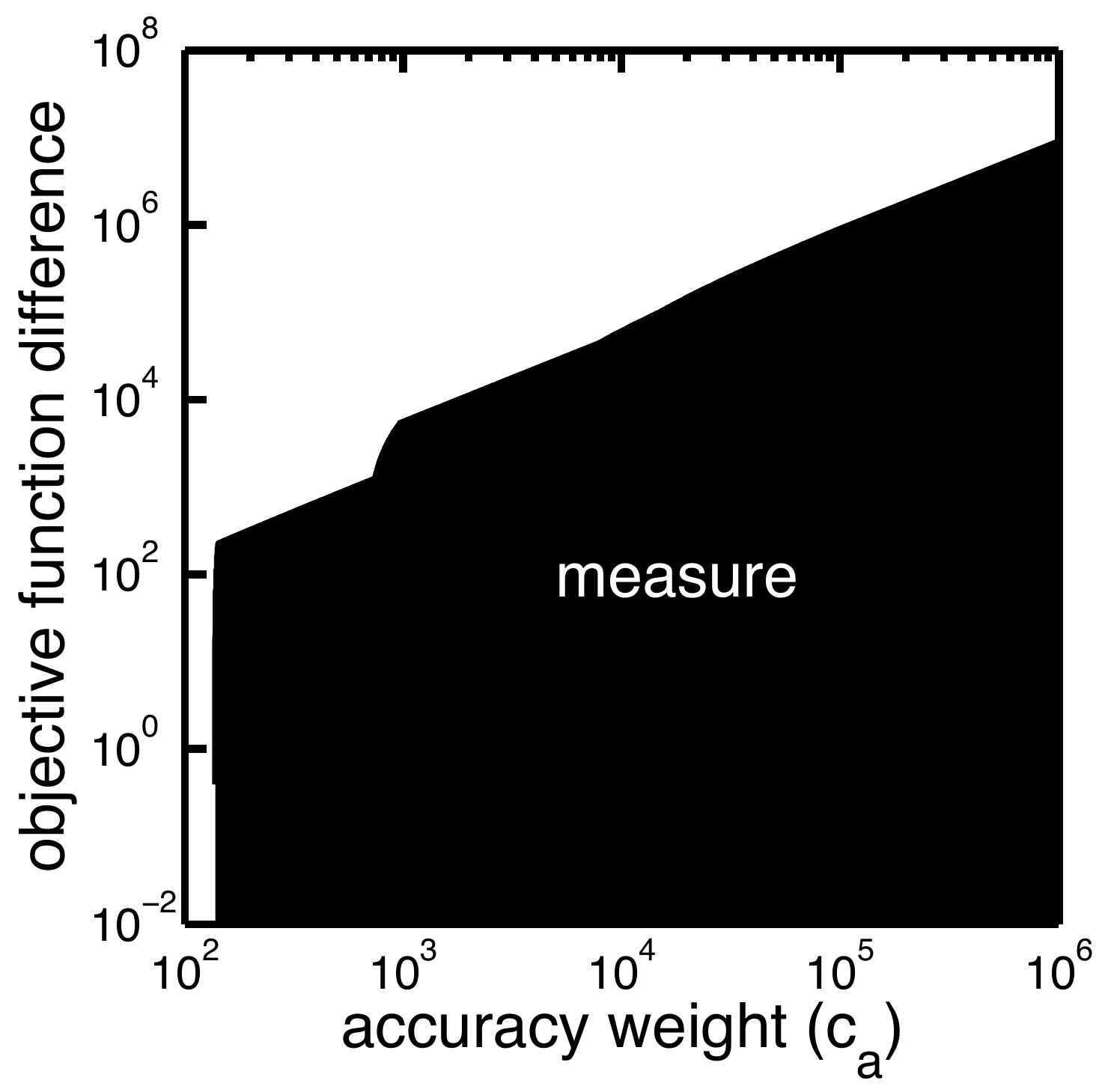}
\caption{{\bf The measurement cost.} The difference in objective functions between using free-living fungi and algae and not using them is shown as a function of the accuracy weight $c_a$ (assuming $c_c=1$). If the measurement cost in the objective function ($c_m M$) is within the black area then measuring free-living fungi and algae will result in a lower objective function. Since the free-living species increase the accuracy of prediction, higher values of $c_a$ correspond to greater tolerance of measurement costs ($c_m M$). When $c_a$ is low, near $10^2$, it is never worth measuring the data because the computation cost of the full system is too expensive.
}
\label{fig:FigMeasure}
\end{figure}
\par
One practical challenge for implementing SSC in empirical systems is that in order to form the best predictive organisms, we need data for all relevant microstate variables. However, some microstate variables may be more costly to measure that others. For example, in the real-world instantiation for our model, free-living fungi and algae are microscopic and difficult to sample compared to macroscopic lichen communities. This makes including such data practically challenging. We can determine the value of such data within our SSC by incorporating the costs of measuring the microstate variables $x$ into the objective function as $M(x)$ in Equation~\ref{Eq:objmeas} ($c_m$ is a conversion factor).
\begin{equation}
f(n,x) = c_a A(n,x) + c_c C(n,x) + c_m M(x)
\label{Eq:objmeas}
\end{equation}
We consider the effects of which microstate data is available by making the accuracy and computation costs functions of both the macrospace dimension $n$ and the microstate variables $x$. We can then compare the difference between the objective functions of the microstate without the free-living species, i.\,e. $x \setminus \{F_1, A_1, F_2, A_2\}$ ($x_r$) and the full microstate ($x_f$), as shown below. 
\begin{eqnarray}
f(n_r,x_r)-f(n_f,x_f) &= c_a \big( A(n_r,x_r) - A(n_f,x _f) \big) \\
&+ c_c \big( C(n_r,x_r) - C(n_f,x_f) \big) \nonumber \\
& - c_m M(\{F_1, A_1, F_2, A_2\}) \nonumber
\end{eqnarray}
The last term $c_m M(\{F_1, A_1, F_2, A_2\})$ represents the cost of measuring the data. If the difference $f(n_r,x_r)-f(n_f,x_f)$ is positive then it is beneficial to measure and include the free-living species in the SSC.
\par
To determine the objective function of the reduced model, without the free-living species, we fit the $\pi$, $\phi$, and $\rho$ maps again but only using the 8 lichen microstate variables, i.\,e., $x \setminus \{F_1, A_1, F_2, A_2\}$. Since there is less information than before when we used the full 12 microstate variables, the accuracy cost is larger for all macrospace dimensions considered, i.\,e. $A(n,x_r) > A(n,x_f)$. Another consequence of using less of the microstate is that the computation cost is reduced. While the $\phi$ and $\rho$ do not change, the  $\pi$ map is now $n \times 8$ instead of $n \times 12$. Thus, the total computation cost is:
\begin{equation}
O_x \big( 6 t n^2 + n(15-3 t)+ 2t \big).
\end{equation}
\par
\noindent With the values of both $A(n_r,x_r)$ and $C(n_r,x_r)$, we can compute the macrospace dimension that minimizes the objective function for $c_a$ and $c_c$. 
\par
The differences in overall cost between the full, twelve-variable, case and the reduced eight-variable case are shown in Fig.~\ref{fig:FigMeasure} as a function of $c_a$ when $c_c$ is normalized to $1$. When $c_a$ is high, large gains are possible if the experimenter can afford to gather data on the additional variables. However, as $c_a$ decreases (or $c_c$ rises), these additional pieces of information become more expensive to include, until, at a critical point, it becomes more efficient to neglect them---no matter how cheap it might have been to gather the data.

\subsection*{The effects of different ecologies}
In all cases thus far we have solved for the best organisms in order to predict quantities of interest. This results in macrostate individuals that are linear combinations of microstate variables. In some cases, however, it may be useful to compare given $\pi$ maps. This may be of particular use in determining whether one set of macrostate variables is a better predictor of quantities of interest than another set, i.\,e., deciding between candidate ``organisms.'' Here, we compare three candidate $\pi$ maps with different notions of organismality corresponding to the different types of microstate variables: free-living fungi and algae, lichen cells, and lichen communities. Furthermore, we vary two ecological conditions to see what effects they have on determining which type of organism is the best predictor. The first ecological condition we vary is the niche structure. Until now, the free-living species and the lichens have been competing for the same resources, i.\,e. sharing a niche. We vary whether or not the niche is shared, and if it is not then we consider different carrying capacities for free-living fungi, free-living algae, and lichen cells. The second ecological condition we vary is the presence/absence of a predator. In real populations, lichen communities are macroscopic entities that can be preyed upon by larger organisms such as deer. As a consequence, we add a predator to the microstate model that only attacks lichen communities (see Methods: Microstate dynamics). 
\par
\begin{figure*}
\centering
\includegraphics[width=1\linewidth]{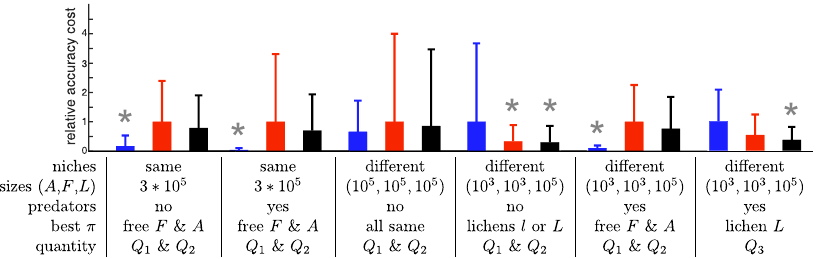}
\caption{{\bf The effects of different microstate model ecologies and quantities of interest.} The mean accuracy cost for three types of organisms: free-living fungi and algae (blue), lichen cells (red), and lichen communities (black) is shown for microstate models incorporating ecological conditions indicated by the table below. Each set of three bars is scaled by the highest average accuracy cost, thus the highest bar in each ecology is the same height. The error bars show the standard deviation over predicting 50 novel trajectories. Gray stars indicate the most accurate ``organism'' according to a Wilcoxon rank sum test with a p-value $<.01$.}
\label{fig:Ecology}
\end{figure*}

For each of the different ecological scenarios, we solve for the $\phi$ and $\rho$ maps that best fit the quantities of interest (as before). Since the three candidate $\pi$ maps are of the same dimension, the computation costs are identical. The key difference therefore rests in the accuracy costs. Figure~\ref{fig:Ecology} shows the effects of ecology on selecting the best organism. When the three types of organisms share the same niche, the free-living species are the best predictors. The presence or absence of a predator has no noticeable effect on this result. Indeed, further explorations suggest that the presence of a predator may reduce the effectiveness of lichen-based maps. This can be seen in the ecologies where free-living fungi, free-living algae, and lichen cells have different niches. If the niches are the same size and there are no predators then all three types of organism are equally predictive. If we reduce the size of the niches for free-living species then we find that the lichens are better predictors (with no significant difference between lichen cells and lichen groups). If we add a predator, then it reduces the number of lichens and keeps them far from carrying capacity---this makes them effectively less predictive of the quantities of interest by making the different niches more equal in size.

Finally, we consider which of the three candidate organisms ($\pi$ maps) is best when we change the quantity of interest. In the microstate models with a predator, both lichen $\pi$ maps became less predictive of $Q_1$ and $Q_2$. But if instead of focusing on the population proportions of $A_1$ and $F_1$ genomes, we turn our attention to the population dynamics of the predator ($Q_3$) then we find a different result. Figure~\ref{fig:Ecology} shows that the lichen groups become significantly better candidate organisms than both lichen cells and free-living species in predicting the population size of the predator. Thus, the choice of the best organism depends on the microstate model ecology and the quantities of interest.

\section*{Discussion}
The organism is a key unit of the biological sciences. Yet it is often defined in a qualitative, subjective fashion and is rarely the subject of quantitative study. We use a framework called State Space Compression \cite{Wolpert2017} that provides a principled way to find ``organisms'' that balance the goals of accurate prediction with the costs of computation and measurement. We apply this framework to a model, inspired by real-world lichen systems, where organismality is particularly at issue. 

We find that even for natural questions---such as the fraction of genotypes of each kind in a population---our results sometimes accord closely with informal notions of organismality, and sometimes yield notions that are less intuitive. As an example of the former, if we allow our system to include only three population-level variables or macrostates, two of them turn out to track the fitter free-living algal and fungal populations very closely. At this scale of analysis, and for these questions, the standard organisms of algae and fungi are in close accord with the best state-space compression. In other cases, however, such as the third macrostate in the $n=3$ case, the individual is a complicated mixture of a number of different variables. While such complicated mixtures might not seem desirable, there is an analogy between SSC and the way scientists visualize data via principle component analysis. It is challenging to picture high dimensional data but by combining variables into often non-intuitive variables, the data can be plotted in a lower-dimensional space where meaningful relationships can be assigned. In other words, what is visually or intuitively obvious need not be the best basic units of prediction or understanding.


Instead of identifying new predictive organisms, it may be desirable to permit only certain types of organisms. We show how our framework can be used to compare given notions of organisms, or $\pi$ maps. Indeed, we chose particular functional forms for the $\pi$, $\phi$, and $\rho$ maps of SSC. Other functions could have been used. This flexibility in the choice of maps has an element of arbitrariness, but it is also the case that for a well-defined scientific application some restriction of the model space is necessary---if only because finding and fitting a truly arbitrary model is impractical, if not impossible. In general, any model-builder must restrict their models to some class; our results highlight the extent to which these deliberate choices help define the best-fit underlying units.

In fact the macrostates in our model gain their meaning from the nonlinear dynamics in the $\phi$ map. To see this, suppose that instead of using (nonlinear) Lotka--Volterra macrostate dynamics we had used purely linear macrostate dynamics. Now whatever macrostate dynamics we use, the individual rows of $\pi$ depend on a choice of basis, which can be accounted for by changing the linear dynamics. So with linear dynamics, it would be only the row-span of $\pi$ that matters, $\phi$ being redundant / irrelevant. In contrast, with Lotka--Volterra dynamics, under a wide range of conditions (which happen to be satisfied by all of our optimal compressions), the only change of bases which preserve the relationship between $\pi$, $\rho$, and $\phi$ are permutations; that is, the set of rows of $\pi$ are well-defined quantities, which we can take to be meaningful. 

The SSC approach to organismality means that the best ``organisms'' depend on the microstate dynamics and the quantities of interest. When we changed the underlying ecology of the microstate model by adding a predator or altering the niche structure, we found different candidate organisms were more predictive of the quantities of interest. Despite this relativism, there may be some general principles that indicate good candidate organisms. For example, the free-living fungi and algae were better candidate organisms in many of the ecological scenarios considered. However, when the lichens were given a much larger, separate niche they became the best predictive organism. This suggests that in ecological scenarios if an entity occupies a separate large niche, then it may be a good predictive organism for some quantities of interest. Furthermore, when we shifted our quantity of interest to the predator population we found that the lichen groups were the best candidate organism. This suggests that if a candidate organism interacts uniquely with other organisms then it might be a good organism for ecological predictions.

State space compression is not the only framework that uses the goals of a task to define its basic units. The use of hidden layers of a deep learning network, for example, can be understood as simultaneously learning the correct coarse-grained units of analysis along with the models that relate them together and allow for their classification and prediction~\cite{mehta2014exact}. The most widely-used machine learning algorithms, however, including deep learning, suffer from interpretability issues. It can be very difficult to understand, or reverse-engineer, what the weights mean (see, e.\,g., Ref.~\cite{landecker}). By contrast, the SSC framework here, by careful choice of $\pi$, $\phi$ and $\rho$, allows us to directly interpret how the individuals are related to more informally-identified components, and to interpret the dynamics of these populations in terms of inhibitory and promoting interactions. 

SSC also highlights the importance of quantifying computation cost; this captures an important trade-off in the scientific endeavor, and is thus an important part of model specification. Our measure depends both on the macrospace dimension $n$ and $t$, the time out to which one wishes to predict. While the specification of computation cost in this toy example is somewhat arbitrary, it is important to note that it does not affect the definitions of the organism (the $\pi$ maps), but only sets the trade-off points where increasing the number of types is preferred. 

By taking this relative, yet formal and quantitative, view of the notion of an organism, our framework has implications for broader issues. The questions we pose here connect directly to issues that are becoming increasingly important as biology becomes more data-rich. It is no longer possible to talk about the natural, ``obvious'' individual units when confronted by, say microbiome data in the form of snippets of DNA gathered without knowledge of phenotype~\cite{o2015backbones}. Evolutionary definitions of the levels of selection are another example of how biologists are coming to re-examine the ways in which we have chosen to define our basic units---whether it be at the level of the gene, the spatially-bounded gamete-producing agent, or the group~\cite{nowak2010evolution}. Finally, we expect that finding the best decomposition of a system's microscopic description into a simpler set of coarse-grained variables is common problem for living organisms. Any organism in an ecosystem must model and predict their own environments just as much as the observant scientist. The general principles that we find for the scientific notion of organismality might apply just as well to the basic categories a predator or competitor might use to map and understand their world.



\section*{Methods}
\subsubsection*{Microstate dynamics}
At each time step of the simulation cells can reproduce, move, lichenize, and die (see SI for the microstate model code and the parameter values). For reproduction, each cell type has a characteristic probability of reproduction, e.\,g. $A_1$ reproduces with probability $p_{A_1}$, $A_2$ reproduces with probability $p_{A_2}$, etc. While algal and fungal reproduction simply results in an additional algal or fungal cell, lichen reproduction has an additional component. Typically, lichens are large multicellular  communities and, as such, reproduction of a single cell (algal-fungal pair) may either contribute to the growth of the existing community (called lichen ``growth'') or lead to formation of a new community (called lichen ``reproduction''). For simplicity, we ignore sexual reproduction of the fungal partners in the lichen. Thus, when a lichen cell reproduces there is a certain probability it will form a new lichen community ($p_{c}$). If a lichen ``grows'' then the new cell is placed a small distance $d_s$ from the parent cell. If, instead, the lichen ``reproduces'' then the new cell is placed a larger distance away $d_l$. The reason for the distinction between distances is that lichen reproduction in the wild occurs via dispersal of lichen fragments which is not the case with lichen growth. In our model, dispersal upon reproduction is the only way that lichen cells move. Free-living fungal and algal cells, however, move according to a 2-D random walk with step size $d_l$. In the simulations, movement occurs after the reproduction step. 
\par
Following movement of free-living algae and fungi, there is the possibility of lichenization. For each free-living alga, we calculate the closest free-living fungus. If the pair are within a certain distance of one another, $d_c$, there is an opportunity to form a lichen. Lichen formation occurs according to a probability $p_l$. If a lichen is formed, then a lichen ``cell'' replaces the algal cell and the fungal cell.  The new lichen cell also forms a new lichen group. Thus, lichenization and lichen reproduction produce the same result: a single algal-fungal pair lichen, forming a new lichen group. 
\par
The last phase in each time step is death. We assume the existence of carrying capacities that limit the total number of algal, fungal, and lichen cells. Initially, we assume that the carrying capacities are shared, which corresponds to the cell types occupying the same ecological niche with a carrying capacity of $3 \times 10^5$. We relax this assumption in the section where we consider different niche structures. If the number of cells is below the carrying capacity, death occurs according to some probability depending on the cell type. If, however, the number of cells is above carrying capacity (as a consequence of reproduction being the first phase of each time step in the simulation), then we restore the population to the carrying capacity by randomly picking cells for death weighted by their characteristic probability of death. After we remove the dead cells, the simulation proceeds to the next time step.
\par
In simulations with predators, we consider a form of death in which a predator may eat an entire lichen community. The predator is similar to some deer that eat large lichen communities but not free-living algae and fungi or lichen communities that are too small to be visible. The probability that a predator eats a lichen community depends on both the number of predators and the size of the lichen community, i.\,e., how noticeable it is. We assume that the probability that this occurs is a product of a hill function of the lichen community size and the number of predators. If a predator eats a lichen community then all cells belonging to that community die. Furthermore, the number of predators increases or decreases depending on the number of lichen cells they eat. Thus, when large communities with many cells are eaten they result in more predators. We track the dynamics of the predators in conjunction with the other microstate variables. For such simulations, the $\phi$ map includes a differential equation for predators and a set of predator interaction terms in all other differential equations. The initial number of predators is mapped into the macrospace using the identity map. As a consequence the number of predators does not factor into fitting the $\pi$ or $\rho$ map. 
\par
The microstate dynamics are simulated for 100 sets of initial conditions. Each initial condition is chosen from a Sobol sequence scaled to 100 such that each cell type begins the simulation with between 1 and 100 cells. Cells are scattered randomly in a 2D square grid of width 100. We simulate the population dynamics for 500 time steps and remove the first 50 time steps to allow populations to expand. If the simulation includes predators, they are not added until after 50 time steps. 
\par
\subsubsection*{Fitting the maps}
We fit the three maps ($\pi$,$\phi$,$\rho$) through a multi-step process because the Lotka--Volterra equations are poorly conditioned for many parameter choices. First, we get an estimate of a good $\pi$ and $\rho$. We ignore $\phi$ and find a pair ($\pi$, $\rho$) that minimizes the Euclidean distance between $\rho(\pi(x_t))$ and $O(x_t))$ for all $t$. This is done through a combination of linear algebra and a least-squares fitting routine ({\tt lsqnonlin} in MATLAB with a Levenberg--Marquardt algorithm). 

With an estimate for ($\pi$, $\rho$), we then hold them fixed and fit $\phi$. We get an estimate for $\phi$ by using the technique in \cite{Stein:2013dl} that takes the log transform of the Lotka--Volterra system and uses linear algebra to fit the parameters assuming that each time step in Lotka--Volterra is weighted evenly. While this helps to give an initial estimate of $\phi$ parameters, they are not optimal because early mistakes in Lotka--Volterra compound as a result of iterating the differential equations. Thus, we apply the same least-squares fitting routine as before while holding $\pi$ and $\rho$ fixed. 
\par
Finally, we use the fits for $\pi$, $\phi$, and $\rho$ as initial conditions for the least-squares fitting routine ({\tt lsqnonlin} in MATLAB with a Levenberg--Marquardt algorithm) to fit the three maps in combination. We fit for $10^5$ evaluations or until a local optimum is reached. The entire sequence of map fittings is repeated 25 times to find the best possible set of parameters.

\subsubsection*{Determining the computation cost}
In the functional forms of the maps we consider there are four basic types of operations: 1. scalar addition, 2. scalar division, 3. scalar multiplication, and 4. matrix multiplication. We use $O_a$, $O_d$, and $O_m$ to represent the cost of a single scalar addition, division, and multiplication operation, respectively. For matrix multiplication we use $O_M(k_1 \times k_2, k_2 \times k_3)$ to represent the cost of multiplying a $k_1 \times k_2$ matrix by a $k_2 \times k_3$ matrix. We assume that the computation cost is to be determined per trajectory, where a trajectory is of length $t$. (If there are $k$ trajectories, the computation cost need only be multiplied by $k$).
\par
Although in principle one could use the optimal worst-case algorithms for these operations, as is standard in computational complexity \cite{CLRS, BCS}, here we instead use common upper bounds that are more reflective of current real-world implementations of the operations to be performed. In practice, we expect that most reasonable estimates of the computation time---either real-world wall-clock or theoretically optimal---would yield roughly similar results, though of course it is possible that disruptive changes to our understanding of the notion of organism could occur given large improvements in technology and algorithms.

Since $\pi$ is a linear map that transforms the initial values of the microstate variables (a 12-dimensional vector) into the macrostate variables (an $n$-dimensional vector, where $n$ is the size of the macrospace) its computation is simply a single matrix multiplication step.
\begin{equation}
C_{\pi} = O_{M}(n \times 12,12 \times 1)
\end{equation}
\par
The $\phi$ map iterates the Lotka--Volterra equations for $t$ time steps over the trajectory. We assume that a simple Euler method is used to solve the differential equations and $dt=1$ (forgoing the need for multiplication). Because Lotka--Volterra equations have both linear and quadratic terms, there are both scalar and matrix multiplication operations in addition to the a scalar addition operator that updates the solution in time. The resulting computation cost is shown below.
\begin{equation}
C_{\phi} = t \big( n O_{a} + n O_{m}+ O_M (n \times n,n \times 1) \big) 
\end{equation}
\par
Finally, the $\rho$ map transforms the macrostate variables into two ratios of linear combinations. The computation cost is shown below.
\begin{equation}
C_{\rho} =  t \big(4 O_M ( 1 \times n,n \times 1) + 2 O_d \big) 
\end{equation}
\par
The computation costs of the various operations will depend on the algorithm and computational tools. Here, we make a few simplifying assumptions to derive a computation cost in terms of the trajectory length $t$ and the number of macrostates $n$.  First we assume that the cost to perform $O_a$, $O_d$, and $O_m$ are the same (call it $O_x$). Second, we assume that a straightforward brute-force method is used for the matrix multiplication such that the cost of $O_M(k_1 \times k_2, k_2 \times k_3)$ is $k_1 k_2 k_3 O_m + k_1 (k_2-1) k_3 O_a$. With these two assumptions the total computation cost is shown in Eq.~\ref{Eq:totcost}.
\begin{equation}
 C_\pi+C_\phi+C_\rho = O_x \big( 6 t n^2 + n(23-3 t)+ 2t \big)
 \label{Eq:totcost}
\end{equation}
\par

\subsection*{Acknowledgments}
DW and EL thank the Templeton World Charity Foundation for funding under grant TWCF0079/AB47. JG and EL thank the SFI Omidyar fellowship for funding. All authors thank the Santa Fe Institute for generous support.

\bibliographystyle{unsrt}
\bibliography{Lichen}

\end{document}